\documentclass[12pt]{article}
\let\section=\subsection     \let\subsection=\subsubsection                
\usepackage[dvips]{graphics,graphicx}
\begin{document}
\begin{center}
 {\bf PION INTERFEROMETRY FROM A RELATIVISTIC FLUID
WITH A FIRST-ORDER PHASE TRANSITION
IN CERN-SPS 158 GeV/A Pb+Pb COLLISIONS}\\[5mm]
 K.~MORITA, S.~MUROYA$^\dagger$, H.~NAKAMURA and C.~NONAKA$^\ddagger$ \\[5mm]
 {\small \it  Department of Physics, Waseda University \\
 169-8555 Tokyo, Japan \\[2mm]}
 {\small \it $^\dagger$Tokuyama Women's College \\ Tokuyama, 745-8511
 Yamaguchi, Japan \\[2mm]}  
 {\small \it $^\ddagger$Department of Physics, Hiroshima University \\
 Higashi-Hiroshima, 739-8526 Hiroshima, Japan \\[3mm]}
\end{center}
\setlength{\unitlength}{1cm}
\begin{picture}(0,0)(0,0)
 \put(13.93,8.7){WU-HEP-00-1}
 \put(14.63,8.2){TWC-00-2}
\end{picture}

\begin{abstract}\noindent
 We investigate pion source sizes through the
 Yano-Koonin-Podgoretski\u{\i} (YKP) parametrization for the Hanbury-Brown
 Twiss (HBT) effect in the CERN-SPS 158 GeV/A central collisions. We
 calculate two-particle correlation functions numerically based on a
 (3+1)-dimensional relativistic hydrodynamics with a first order phase
 transition and analyze the pair momentum dependence of the HBT radii
 extracted from the YKP parametrization in detail. We find that even in the
 case of a first order phase transition, expansion and the surface dominant
 freeze-out make the source in the hydrodynamical model opaque
 significantly. Consequently, the interpretation of the temporal radius
 parameter as the time duration becomes unavailable for the hydrodynamical
 model. 
\end{abstract}

 Pion interferometry is one of the promising tools to investigate the hot
 and dense matter created in the relativistic heavy ion collisions
 \cite{review}. It is well known as Hanbury-Brown Twiss (HBT) effect that we
 can obtain a size of the source from a two-particle correlation function
 because of the symmetry of the wave function of the emitted bosons.  In the
 experimental analyses, the size of the source is obtained as a fitting
 parameter in the Gaussian fit to the correlation function and its physical
 meaning depends on the Gaussian fitting function. For central collisions,
 Yano-Koonin-Podgoretski\u{\i} (YKP) parametrization is used as one of the
 fitting functions. In the YKP parametrization, three spatial parameters are
 interpreted directly as the transverse (perpendicular to the collision
 axis), longitudinal (parallel to the collision axis) and temporal (emission
 duration) extents, respectively \cite{ykp}. In the case of relativistic
 heavy ion collisions, the reaction is highly dynamical. The collective
 expansion, which contains some informations of an equation of state of the
 matter, takes place naturally. For example, if there exists a first order
 phase transition between quark-gluon plasma (QGP) and hadronic gas, enlarged
 time duration due to the existence of the mixed phase may be observed
 \cite{pratt}. However, the collective expansion makes the meaning of the
 size ambigious. Therefore, we investigate the sizes in the YKP
 parametrization based on a hydrodynamical model with a first order phase
 transition \cite{paper}.

 Assuming a completely chaotic source for simplicity, the two-particle
 correlation function is given as 
 \begin{equation}
  C({\bf k_1,k_2})
   =1+\frac{|I({\bf k_1,k_2})|^2}{(dN/d^3{\bf k_1})(dN/d^3{\bf k_2})},
   \label{c2}
 \end{equation}
 where ${\bf k_1}$ and ${\bf k_2}$ are momenta of a emitted pair and
 $I({\bf k_1,k_2})$ is the interference term. 

 As usual,  we introduce the relative momentum $q^\mu=k_1^\mu-k_2^\mu$ and
 the average momentum $K^\mu=(k_1^\mu+k_2^\mu)/2$ that satisfy the on-shell
 condition $q_\mu K^\mu=0$. Assuming the cylindrical symmetry, we can take
 $K^\mu=(K^0,K_T,0,K_L)$. Hence, only three components of relative momenta
 are independent. In the YKP parametrization, $q_\bot=\sqrt{q_x^2+q_y^2}$,
 $q_\|=q_z$, and $q^0$ are taken as the independent components. Then the
 fitting function is given as
\begin{eqnarray}
 C(q^\mu,K^\mu)\!\!\! &=&\!\!\! 1+\lambda\exp 
  \left\{ 
   -R_\bot^2(K^\mu)q_\bot^2-R_\|^2(K^\mu)[q_\|^2-(q^0)^2]
\right.\nonumber\\
&& \qquad \qquad \left. -\left[
	 R_0^2(K^\mu)+R_\|^2(K^\mu)
	 \right]
 \left[
  q_\mu u^\mu (K^\mu)
  \right]^2 
  \right\} ,
\label{ykp}
\end{eqnarray}
where $u^\mu$ is a four-velocity which has only longitudinal component
$v(K^\mu)$ interpreted as a expansion velocity and the fitting parameters are
$v(K^\mu)$ and $R_i$ $(i=\bot, \|, 0)$. Because the three size parameters are
invariant under longitudinal boosts, we may use a special frame called YKP
frame, where $v=0$, in the discussion of the meanings of the size
parameters. Introducing the source function as 
\begin{equation}
 S(x^\mu, K^\mu)=\int_\Sigma \frac{U_\mu(x^\prime)d\sigma^\mu(x^\prime)}
  {(2\pi)^3}
  \frac{U_\nu(x^\prime)k^\nu}{\exp(U_\rho(x^\prime)k^\rho/T)-1}
  \delta^4(x-x^\prime), \label{S} 
\end{equation}
and the weighted average 
\begin{equation}
 \langle A(x^\mu) \rangle =\frac{\int d^4x A(x^\mu) S(x^\mu,K^\mu)}
  {\int d^4 x S(x^\mu,K^\mu)},
\end{equation}
the size parameters in the YKP frame are expressed as 
\begin{eqnarray}
 R_\bot^2(K^\mu)&=& (\Delta y)^2,\label{rt}\\
 R_0^2(K^\mu)&=& (\Delta t)^2
  -\frac{2}{\beta_\bot}\langle \tilde{x}\tilde{t} \rangle
  +\frac{1}{\beta_\bot^2}[(\Delta x)^2-(\Delta y)^2],\label{r0}\\
 R_\|^2(K^\mu)&=&(\Delta z)^2
  -\frac{2\beta_\|}{\beta_\bot}\langle \tilde{x}\tilde{z} \rangle
  +\frac{\beta_\|^2}{\beta_\bot^2}[(\Delta x)^2-(\Delta y)^2],\label{rl}
\end{eqnarray}
where $\tilde{x}=x-\langle x \rangle$, $\Delta x=\sqrt{\langle x^2 
\rangle-\langle x \rangle^2}$, $\beta_\bot=K_T/K^0$ and $\beta_\|=K_L/K^0$.
From the expressions above, we can interprete the size parameters as the
extents of the source if the last two terms in Eqs. (\ref{r0}) and (\ref{rl})
are sufficiently small. It has been already shown that the interpretation
holds within a class of thermal source model \cite{ykp}, and the case of a
hydrodynamical model is the point of the present paper.

Figure 1 shows $K_T$ dependence of the pion HBT radii. Close circles stand
for the HBT radii extracted from the pion correlation function (\ref{c2})
through the fitting. Solid lines are space-time extensions calculated from
Eqs. (\ref{rt})-(\ref{rl}), which are expected to agree with the HBT radii
when only thermal pions are considered. Dotted lines stand for the ``source
sizes'' ($\Delta z$ for $R_\|$ and $\Delta t$ for $R_0$). As far as $R_\bot$
and $R_\|$ are concerned, those quantities seem to be consistent and well 
reproduce the NA49 experimental results \cite{na49} (open circles). As for
$R_0$, our results are a few fm smaller than the experimental results in spite
of the existence of the first order phase transition. Furthermore, the time
durations $\Delta t$ do not agree with the $R_0$ at large $K_T$, that means
the failure of the interpretation of $R_0$ as the time duration. As explained
below, this failure comes from an opaque property of the source. 

\vspace*{1cm}
 \begin{minipage}[c]{13cm}
  \baselineskip=12pt
  \begin{center}
   \includegraphics[scale=0.3]{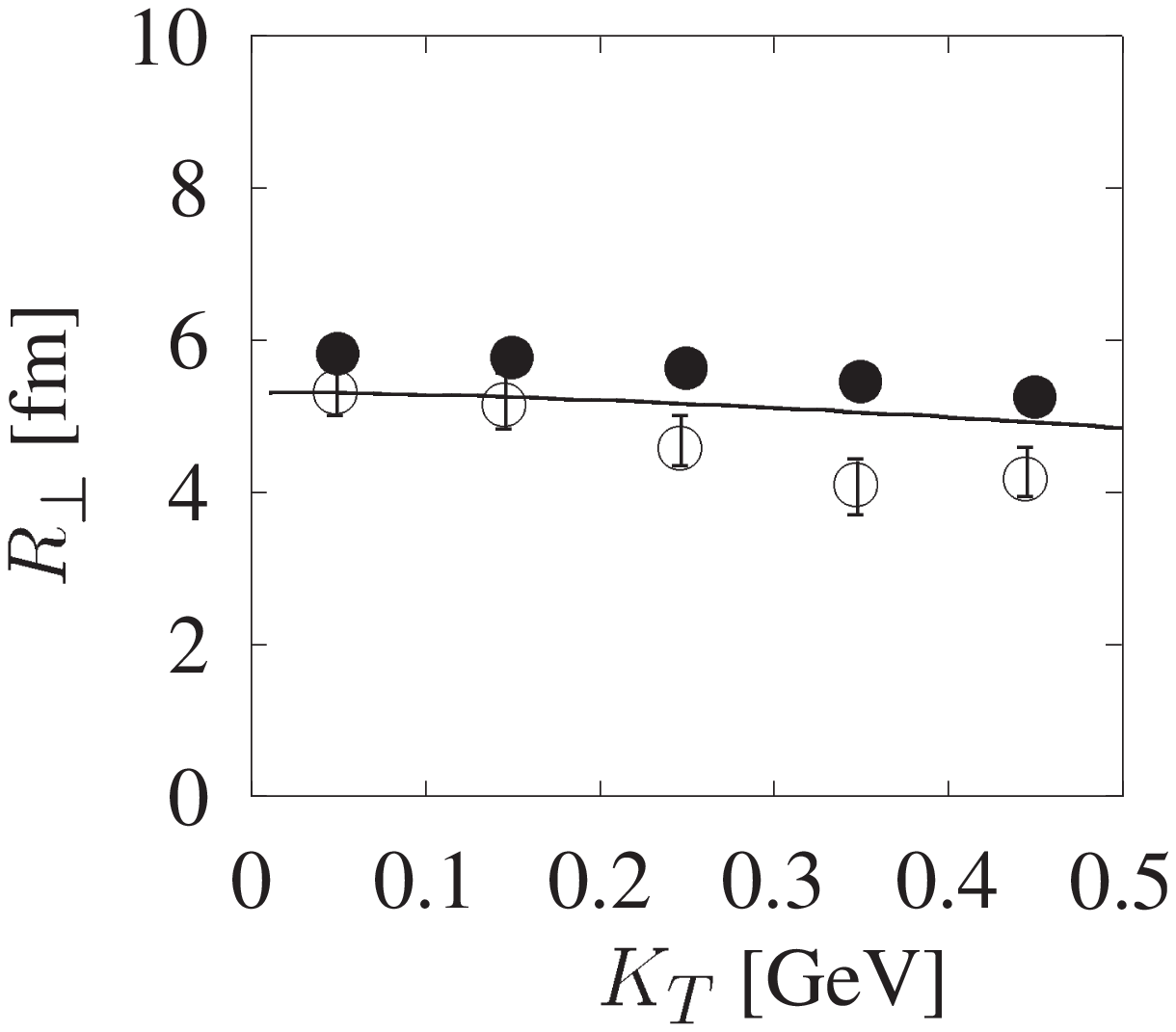}\,
   \includegraphics[scale=0.3]{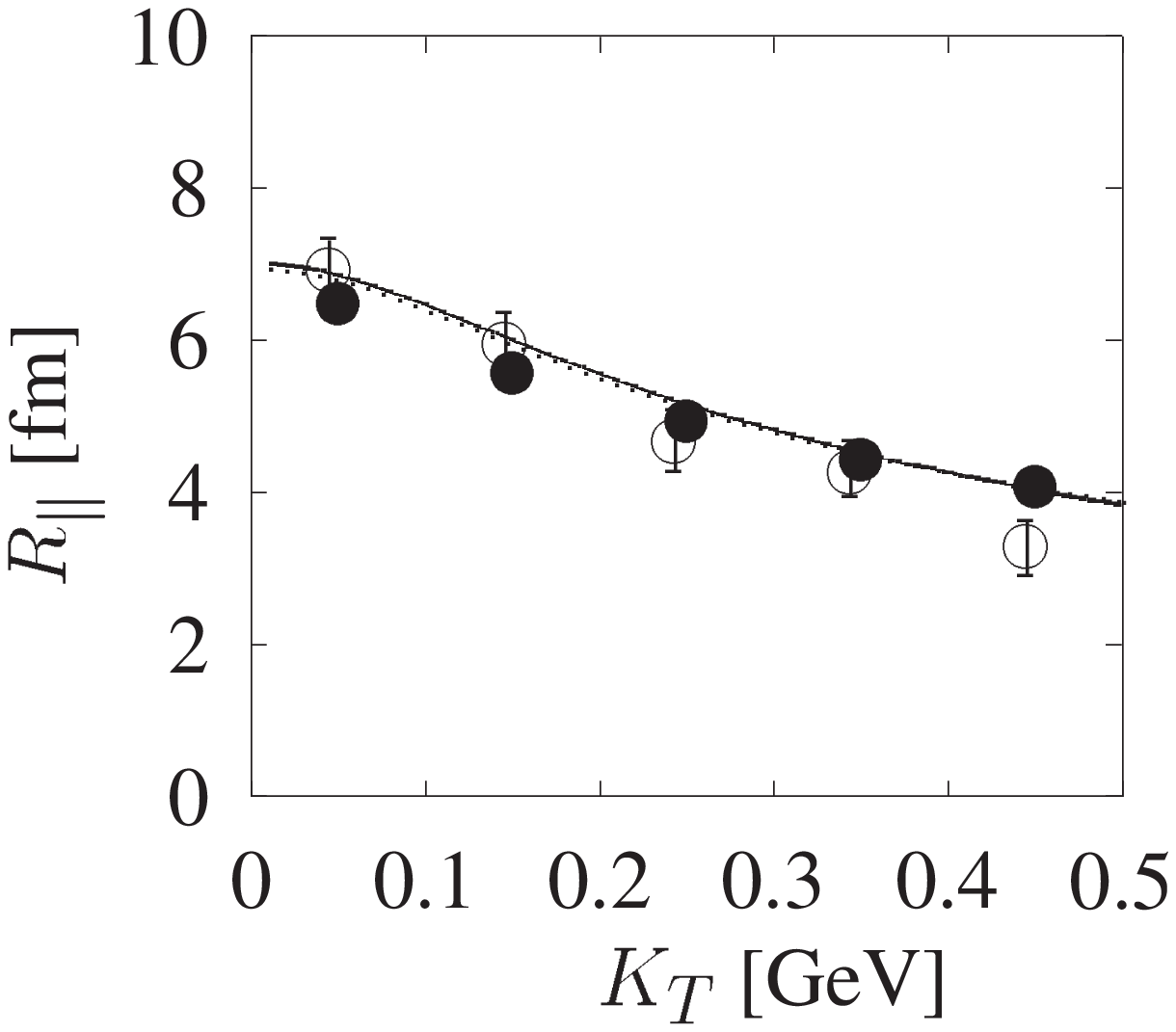}\,
   \includegraphics[scale=0.3]{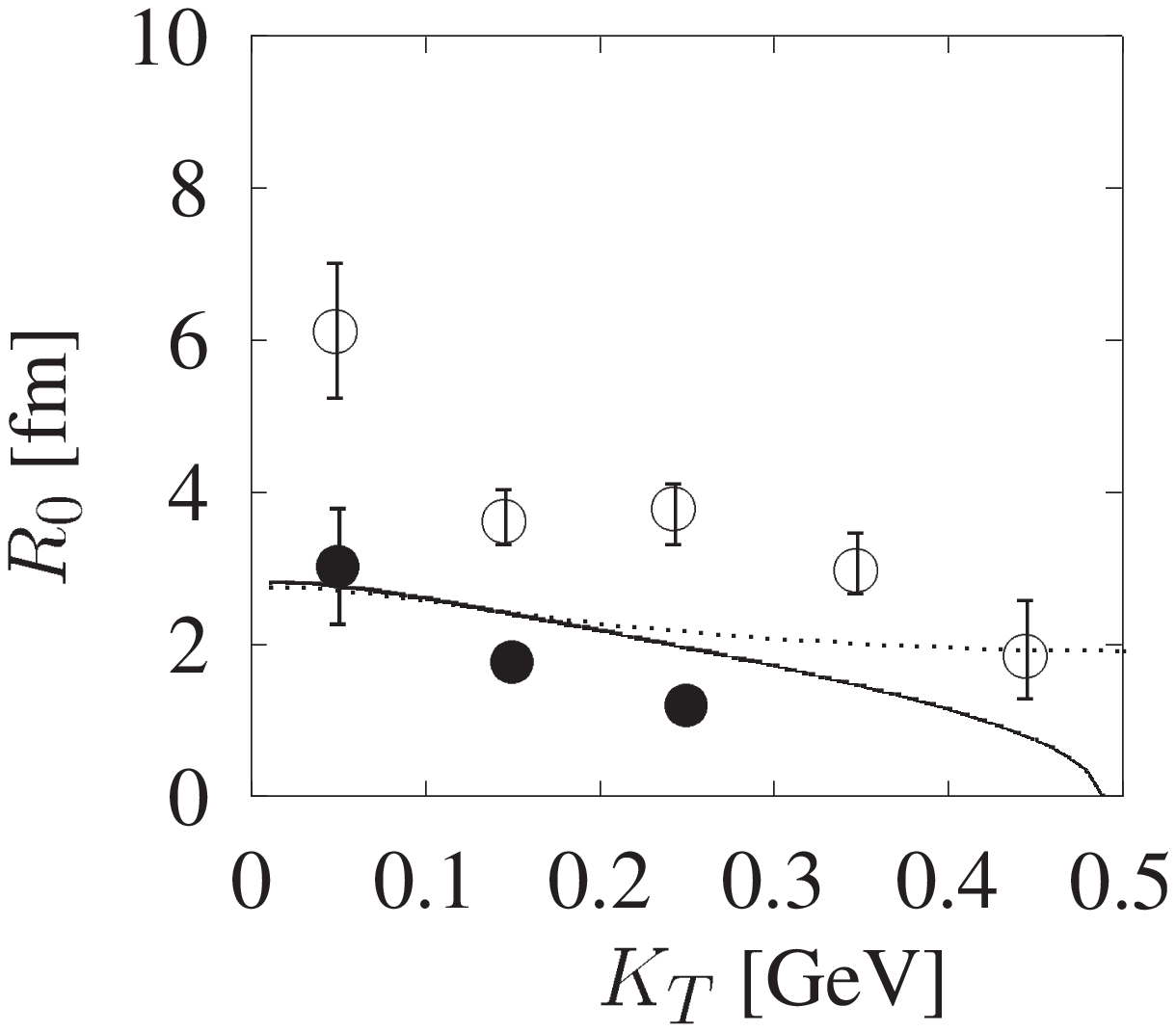}\\
  \end{center}
  {\small Fig.~1 $K_T$ dependence of YKP radii. Open and closed circles show
  the experimental data and HBT radii, respectively. Solid lines stand for
  space-time extensions  (\protect\ref{rt})-(\protect\ref{rl}). Dotted lines
  stand for source sizes. ($\Delta z$ for $R_\|$ and  $\Delta t$ for
  $R_0$.)}
 \end{minipage}\\

The ``opaque'' source \cite{opaque} emits particles dominantly from the
surface and the ``transparent'' source emits from whole region. Figure 2
shows the source function (\ref{S}) projected onto the $x-y$ plane,
$\tilde{S}_T(x,y)=*\int dz dt S(x^\mu,K^\mu)$ being normalized as $\int dx dy
\tilde{S}_T(x,y)=1$. Note that the average momenta of the emitted pions is
$x$ direction in the figure. The left figure shows \textit{clearly} surface
emission; pions are emitted from the crescent region and the source is
thinner in the $x$ direction than in the $y$ direction. This is typical
property of the opaque source and should be appeared in the third and fourth
term in Eq. (\ref{r0}) and the factor $(\Delta x)^2-(\Delta y)^2$ works as a
measure of source opacity. On the other hand, if we neglected the transverse
flow by putting $U^r=0$ by hand, the source function becomes as shown in the
right figure of Fig.~2. In this case, the source function is proportional to
the space-time volume of freeze-out hypersurface. As a consequence of
restoration of the azimuthal symmetry which is shown clearly in the figure,
the above measure of source opacity vanishes. Although the particle emission
takes place almostly from the thin surface, the source is not opaque in this
sence. When the transverse flow exists, the source function is deformed by
thermal Boltzmann factor $\exp(K_T \sinh Y_T \cos \phi /T_f)$ with $Y_T$
almost proportional to $r$. Consequently, the number of emitted particles
increases in the region where $x>0$ (i.e., $\cos \phi >0$) and decreases
in the region where $x<0$ ($\cos \phi <0$). This flow effect deforms the
surface dominant volume distribution (right figure) to the crescent shape
(left figure). 

\vspace*{1cm}
  \begin{minipage}[c]{13cm}
   \baselineskip=12pt
   \begin{center}
	\includegraphics[scale=0.6]{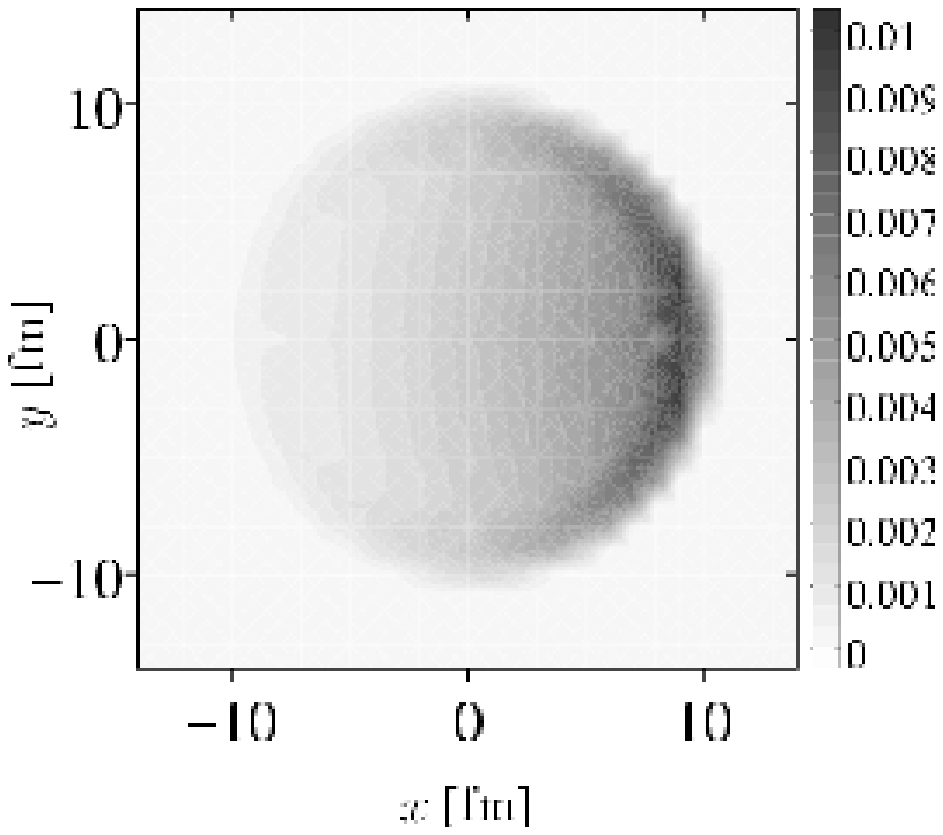}\,
	\includegraphics[scale=0.6]{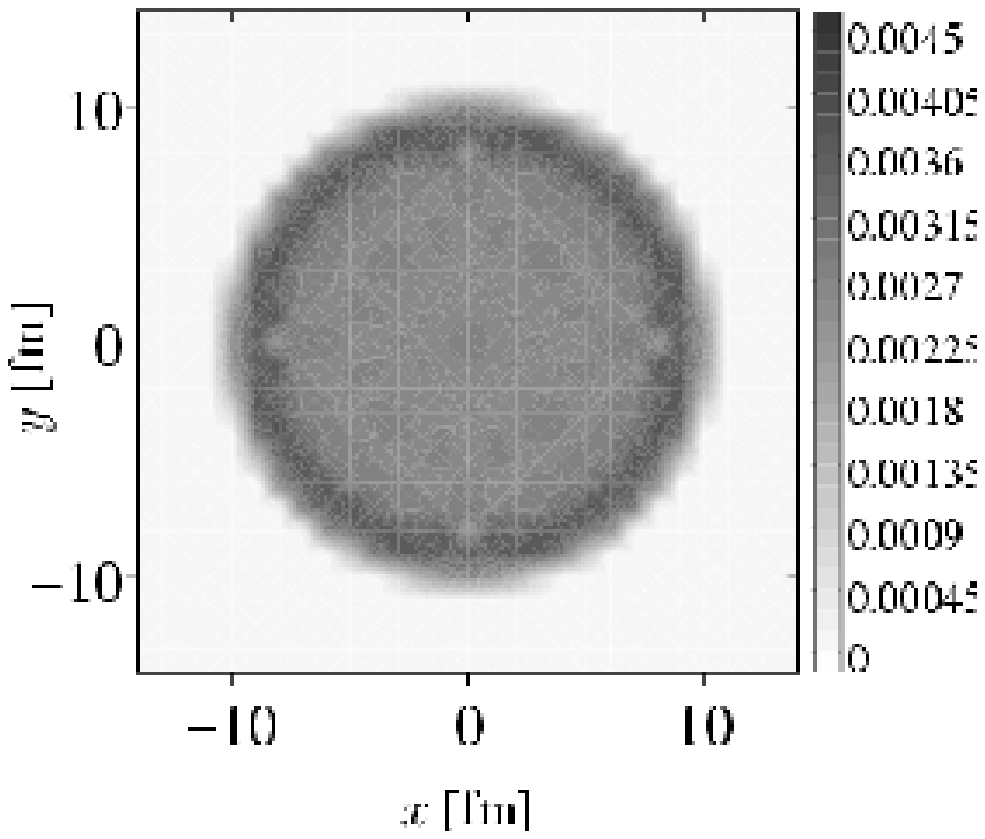}
   \end{center}
  {\small Fig.~2 The source functions as functions of transverse coordinates
   $x$ and $y$ for $K_T=450$ MeV and $Y_{\pi\pi}=4.15$
   ($Y_{\pi\pi}=\frac{1}{2}\ln \frac{K^0+K_L}{K^0-K_L}$). The left figure
   denotes the source function with the transverse flow. The right figure
   denotes the source function \textit{without} transverse flow.}
  \end{minipage}\\

In summary, we analyze $K_T$ dependence of source parameters of the YKP
parametrization based on the relativistic hydrodynamics for the CERN-SPS 158
GeV/A Pb+Pb collisions. We obtain the results almostly consistent with the
experiment. However, the temporal source parameter, $R_0$, shows the behavior
different from the experiment. The source opacity makes the interpretation of
$R_0$ as the time duration hard. We found that the source opacity was
caused by the transverse flow and the characteristics of the freeze-out
hypersurface, the surface dominant freeze-out. The deviations of our results
from the experiment would be improved by including the resonance decay and
other effects.

The authors are indebted to Professor I.~Ohba and Professor H.~Nakazato for
their helpful comments. This work was partially supported by a Grant-in-Aid
for Science Research, Ministry of Education, Science and Culture, Japan
(Grant No. 09740221) and Waseda University Media Network Center.

\end{document}